\documentclass[preprint]{aastex}

\shorttitle{The Old Open Cluster Be 17}
\shortauthors{Krusberg \& Chaboyer}

\begin{document}

\title{UBVI CCD Photometry of the Old Open Cluster Berkeley 17}
\author{Zosia A. C. Krusberg and Brian Chaboyer}
\affil{Department of Physics and Astronomy, 6127 Wilder Lab, 
Dartmouth College, Hanover, NH 03755}

\begin{abstract}

Photometric UBVI CCD photometry is presented for NGC 188 and Berkeley
17.  Color-magnitude diagrams (CMDs) are constructed and reach well
past the main-sequence turn-off for both clusters.  Cluster ages are
determined by means of isochrone fitting to the cluster CMDs.  These
fits are constrained to agree with spectroscopic metallicity and
reddening estimates.  Cluster ages are determined to be $7.0 \pm
0.5\,$ Gyr for NGC 188, and $10.0 \pm 1.0\,$Gyr for Berkeley 17, where
the errors refer to uncertainties in the relative age determinations.
These ages are compared to the ages of relatively metal-rich inner
halo/thick disk globular clusters and other old open clusters.
Berkeley 17 and NGC 6791 are the oldest open clusters with an age of
10 Gyr.  They are 2 Gyr younger than the thick disk globular clusters.
These results confirm the status of Berkeley 17 as one of the oldest
known open cluster in the Milky Way, and its age provides a lower
limit to the age of the Galactic disk.

\end{abstract}


\section{Introduction}

Galaxy formation and evolution theory remains one of the great
outstanding problems in contemporary astrophysics.  Although
considerable progress has been made in this field over the past
decades, there are many unanswered questions regarding the formation
of galaxies like the Milky Way exists.  Observations within the Milky
Way galaxy can be used to probe galactic evolution.  In particular,
determining the relative ages of the different stellar populations in
the Milky Way --- the halo, thick disk, thin disk, and bulge --- by
dating open and globular stellar clusters provides significant insight
into the chronology of Galaxy formation \citep{liu00,sal04}.

Owing to the great importance of the old open cluster population in
probing the chemical and dynamic evolution of the Galaxy, considerable
effort has been devoted to determining the physical parameters of
these clusters \citep[e.g.][]{phe94,jan94,sco95}.  However, relatively
few of the oldest open clusters have received commensurate attention
in astrometric, photometric, and spectroscopic studies, despite the
great promise they hold in determining the age of the Galactic disk
(notable exceptions include NGC 188 and NGC 6791).

This paper presents accurate photometric UBVI photometry for old open
clusters NGC 188 and Berkeley 17.  A number of excellent studies have
previously been performed for NGC 188 that have accurately determined
its physical parameters \citep{von98, sar99, pla03,mic04,ste04,van04}.  
NGC 188 is used as a fiducial
cluster, and a precise, relative age is obtained for Berkeley 17
relative to NGC 188.  Additionally, by directly comparing it with
previously published data, the NGC 188 photometry serves as an
independent test of the accuracy of the photometric calibration from
the instrumental to the standard system.

Since the first photometric study of NGC 188 by \cite{san62}
demonstrating that it belonged to the oldest open clusters in the
Galactic disk, the cluster has been the subject of numerous studies.
As the highest priority cluster in the WIYN Open Cluster Study,
excellent, multi-color CMDs and proper-motion data have been
published for this cluster \citep{von98,sar99,pla03}.  In their 
UBVRI photometric study, \cite{sar99} found that NGC 188 has an age of
$7.0 \pm 0.5\,$Gyr, a reddening of E(B--V) = 0.09 $\pm$ 0.02, and a
distance modulus of (m--M)$_V$ = 11.44 $\pm$ 0.08.  These values are
in general agreement with other recent photometric studies of the
cluster \citep{cap90,twa89}.  \cite{ste04} obtained new data on NGC
188 and did a comprehensive review of existing data in the literature
resulting in a large, homogeneous photometric database.  These data were
used by \cite{van04} to determine an age of $6.8\pm 0.7\,$Gyr for NGC
188.

Berkeley 17 was discovered by \cite{set62}, and is located at a low
Galactic latitude in the direction of the Galactic anticenter.  As a
consequence of the cluster's location as well as its large distance,
the field in the direction of Berkeley 17 is highly reddened and
greatly contaminated by field stars.  Since the extensive open cluster
study by \cite{phe94}, a general agreement has prevailed that Berkeley
17 is indeed the oldest known open cluster in the Galaxy.  Soon after
the Phelps et al.~study, \cite{kal94} performed the first BVI
photometry of the cluster, and established that Berkeley 17 is as old
or somewhat older than the previously oldest known open cluster NGC
6791, based on comparisons of the morphologies of the two clusters'
CMDs.  However, the Kaluzny study was hampered by poor weather,
thereby preventing absolute determinations of Berkeley 17's physical
parameters.

The first age determinations for Be 17 based on isochrone fitting were
made by \cite{phe97} who found an age of 10--13 Gyr, a metallicity of
--0.30 $<$ [Fe/H] $<$ 0.00, a reddening of 0.52 $<$ E(B--V) $<$ 0.68
and 0.61 $<$ E(V--I) $<$ 0.71, and a distance modulus of (m--M)$_V$ =
14.05 $\pm$ 0.25.  Although other attempts at establishing the age of
Berkeley 17 have been carried out \citep{car99b,sal04}, no further
observational data has been published for the cluster.

Section \ref{section2} describes the observations
and the data reduction process, and \S \ref{n188photometry}
presents cluster CMDs.  Section \ref{section4} describes our stellar
models, isochrone fits, and the age determinations of the clusters.  These
ages are compared to the ages of other old open clusters and relatively 
metal-rich globular clusters in \S \ref{compare}.
Section \ref{section5} evaluates these results in the
context of the formation of the Milky Way.

\section{CCD Photometry} \label{section2}

\subsection{Observations}

Observations were made at the MDM Observatory 1.3m McGraw-Hill
telescope for thirteen nights between September 7 and 19, 2004.  The
Templeton camera, a thinned, backside-illuminated SITE 1024 $\times$
1024 pixel CCD, was used, producing an image scale of 0.50 arcsec per
pixel and a field of view of 8.5 $\times$ 8.5 arcmin.  The
observations are summarized in Table \ref{table3}.

Preliminary processing of the data was carried out within the
IRAF\footnote{IRAF is distributed by the National Optical Astronomy
Observatories, which is operated by the Association of Universities
for Research in Astronomy, Inc., under cooperative agreement with the
National Science Foundation.} data reduction software and consisted of
standard zero-level correction and flat fielding.  For the zero-level
correction, between 15 and 20 bias frames were obtained throughout
each night.  Twilight and dawn flats were acquired every evening and
used for flat fielding.

Four standard star fields from \cite{lan92} --- PG1633, PG2213, and
two fields in L95 --- were observed throughout the potentially
photometric nights.  These fields consisted of a total of 22 stars
with a color range of $-0.22 \leq (B-V) \leq 2.00$ and were observed
at airmasses ranging from 1.1 to 1.8 to ensure accurate determination
of the photometric extinction and color terms.

\subsection{Standard Star Frame Reduction} \label{transf}

Aperture photometry was completed on the standard star frames using
the APPHOT package within IRAF\@.  The transformation equations
between the instrumental and standard systems were established using
the PHOTCAL package; these equations contain a zero point, an
extinction term, and a linear color term, and are given by:

\begin{center}

$u=U+u_1+u_2X+u_3(U-B)$\\
$b=B+b_1+b_2X+b_3(B-V)$\\
$v=V+v_1+v_2X+v_3(B-V)$\\
$i=I+i_1+i_2X+i_3(V-I)$\\

\end{center}

\noindent where $u$, $b$, $v$, and $i$ are the instrumental
magnitudes, $U$, $B$, $V$, and $I$ are the standard magnitudes, and
$X$ is the airmass.  The color term coefficients represent a weighted
mean average from the three photometric nights (September 10, 14, and
15), and were found to be $u_3=-0.025$, $b_3=-0.040$, $v_3=-0.014$,
and $i_3=0.014$.  The zero-point and extinction terms, however, showed
sufficient night-to-night variation to be determined separately for
each night, assuming the average color terms.  No trends were observed
in the residuals to the fits with respect to time, airmass, or color.
Typical rms deviations of the fits to the standard values were 0.10
mag for U, and 0.02 mag for B, V, and I.

\subsection{Cluster Frame Reduction}

Point-spread function (PSF) photometry of the cluster frames was
performed with the DAOPHOT II \citep{ste92} photometry package.  An
average of 80 uncrowded stars from each frame were selected to
construct a PSF\@.  The PSF photometry was then carried out with the
ALLSTAR package by fitting the final PSF to detected profiles on each
frame.  The subtracted frames were then examined to ensure they
contained only saturated stars, galaxies, and chip defects.  Stars
with magnitude errors larger than 0.1 were edited out of the final
photometry file.

Aperture corrections were determined by performing aperture photometry
on 100--150 bright, uncrowded stars in each frame.  The difference
between the magnitudes given by the PSF and aperture photometry
constituted the aperture correction, which was then applied to the
original PSF photometry.  Typical values for the aperture correction
ranged between 0.1 and 0.2, and the values showed no trends in
variation across the chip.

\section{Color-Magnitude Diagrams} \label{n188photometry}

\begin{figure}
\epsscale{.75}
\plotone{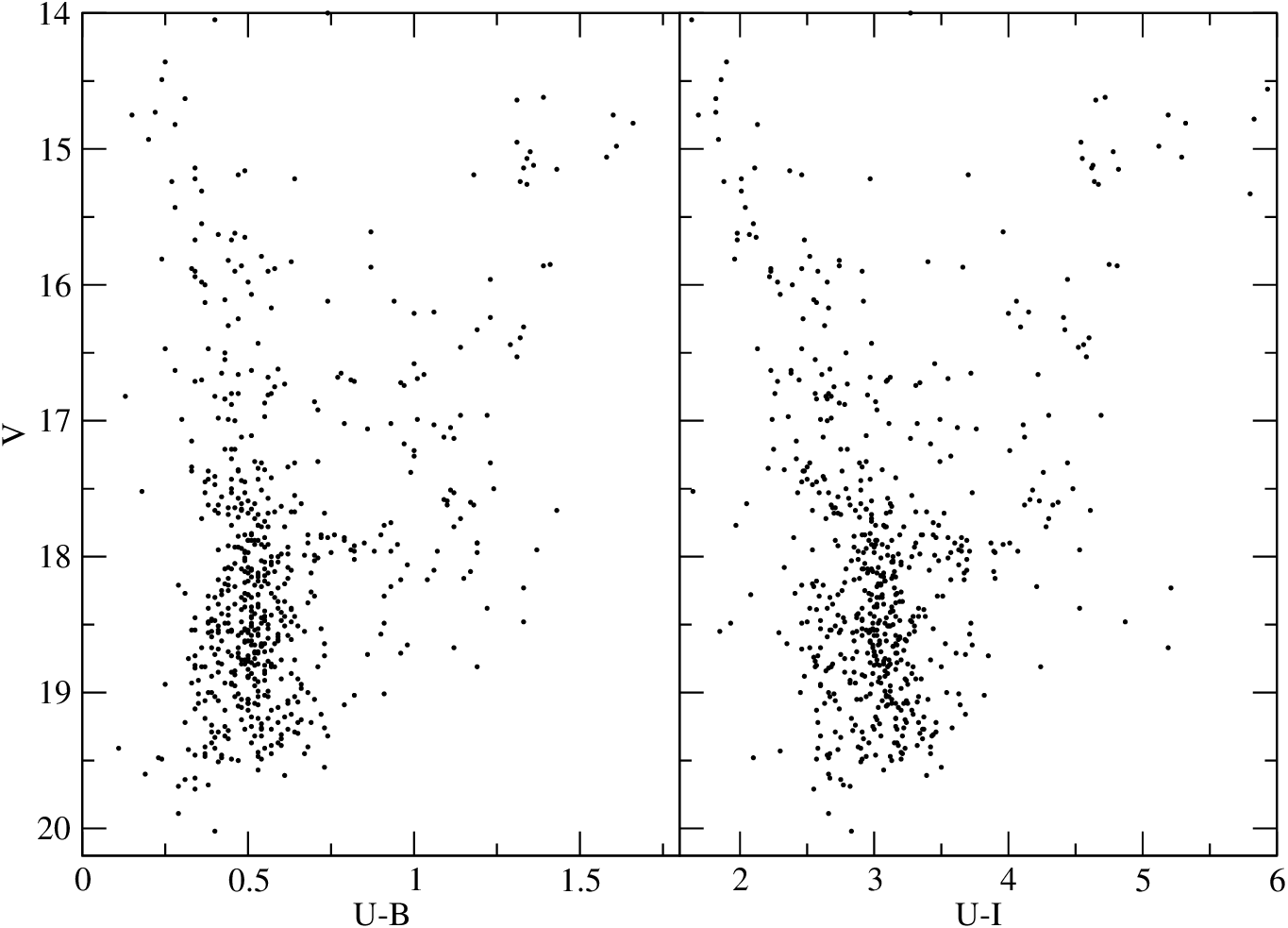}
\caption{Berkeley 17 $(V, U-B)$ and $(V, U-I)$ CMDs.\label{be17ucmd}}
\end{figure}

\begin{figure}
\epsscale{.75}
\plotone{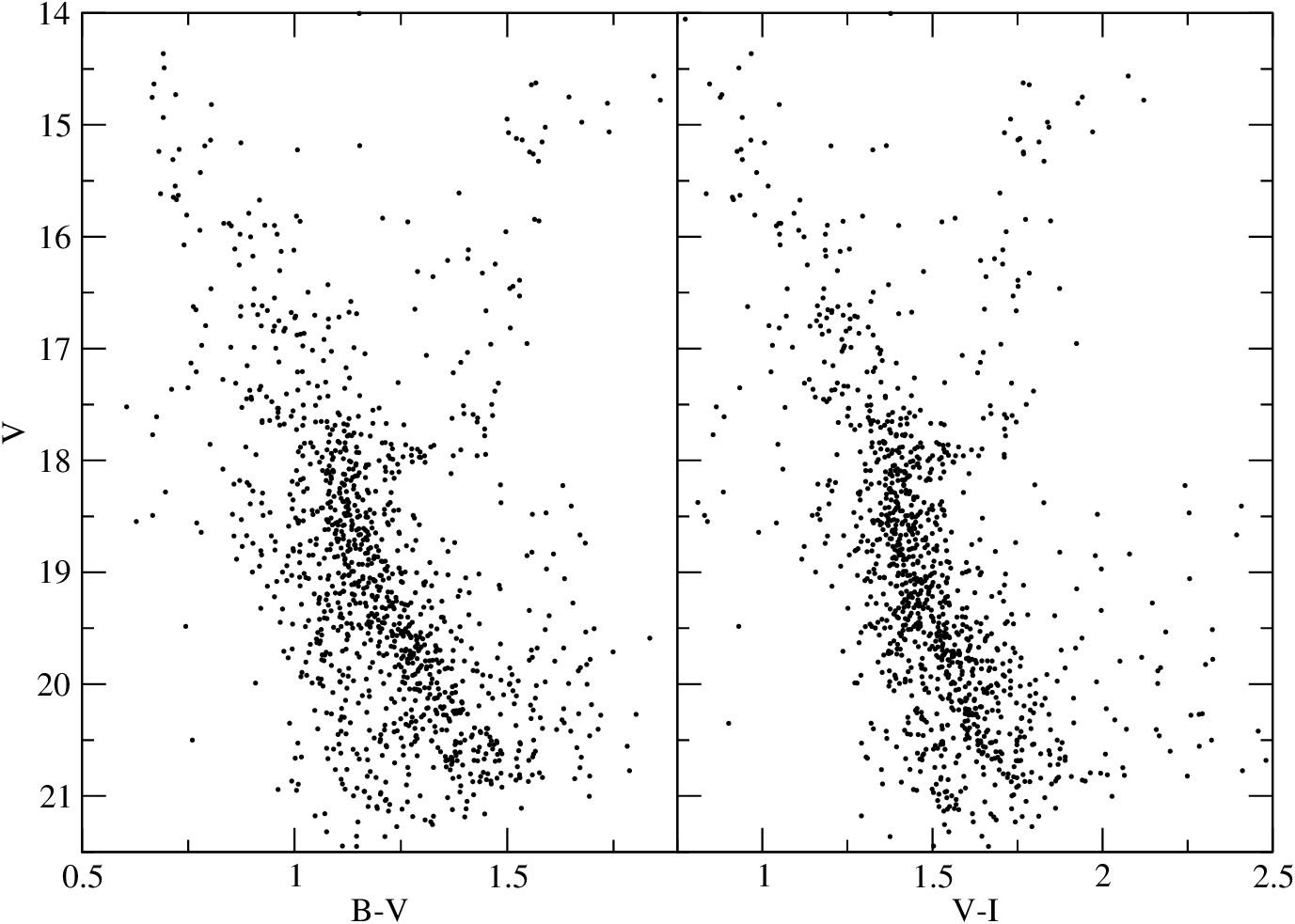}
\caption{Berkeley 17 $(V, B-V)$ and $(V, V-I)$ CMDs.\label{be17cmd}}
\end{figure}

Color-magnitude diagrams (CMDs) for the clusters were constructed
using the standard magnitudes given by the transformation equations
(see Section \ref{transf}).  Stars detected in at least two frames of
a given filter and exposure time were averaged, and were included in
the final CMD if they also appeared in two other filters.  The final
Be 17 cluster CMDs are shown in Figures \ref{be17ucmd}--\ref{be17cmd},
and photometry files are given in Table \ref{n188phottable} for NGC
188 and Table \ref{be17phottable} for Be 17.  Entries of 99.99 in
these tables indicate that a star was not detected in that filter.

The NGC 188 photometry includes only stars within the central region
of the cluster, as the cluster has an estimated angular size of
approximately 1$^{o}$.  The photometry extends around 6 magnitudes
fainter than its main-sequence turnoff at $V\approx 15$.  The CMD
includes 536 stars, and displays a well-defined main-sequence,
subgiant branch, and red giant branch.  A helium-burning clump cannot
be accurately identified.  A binary sequence located 0.75 magnitudes
above the main sequence is detected, most notably in the $(V, V-I)$
CMD. Our photometry covers a much smaller area, and is not quite as deep 
as the photometry presented by \cite{sar99} and \cite{ste04}.  Our
photometry of NGC 188 was obtained so that  a precise, relative comparison 
could be made be the well studied cluster NGC 188 and Be 17.  In addition, 
the NGC 188 can be used to assess the absolute calibration of our 
photometry, via a comparison to other data sets.  

\begin{figure}
\epsscale{.75}
\plotone{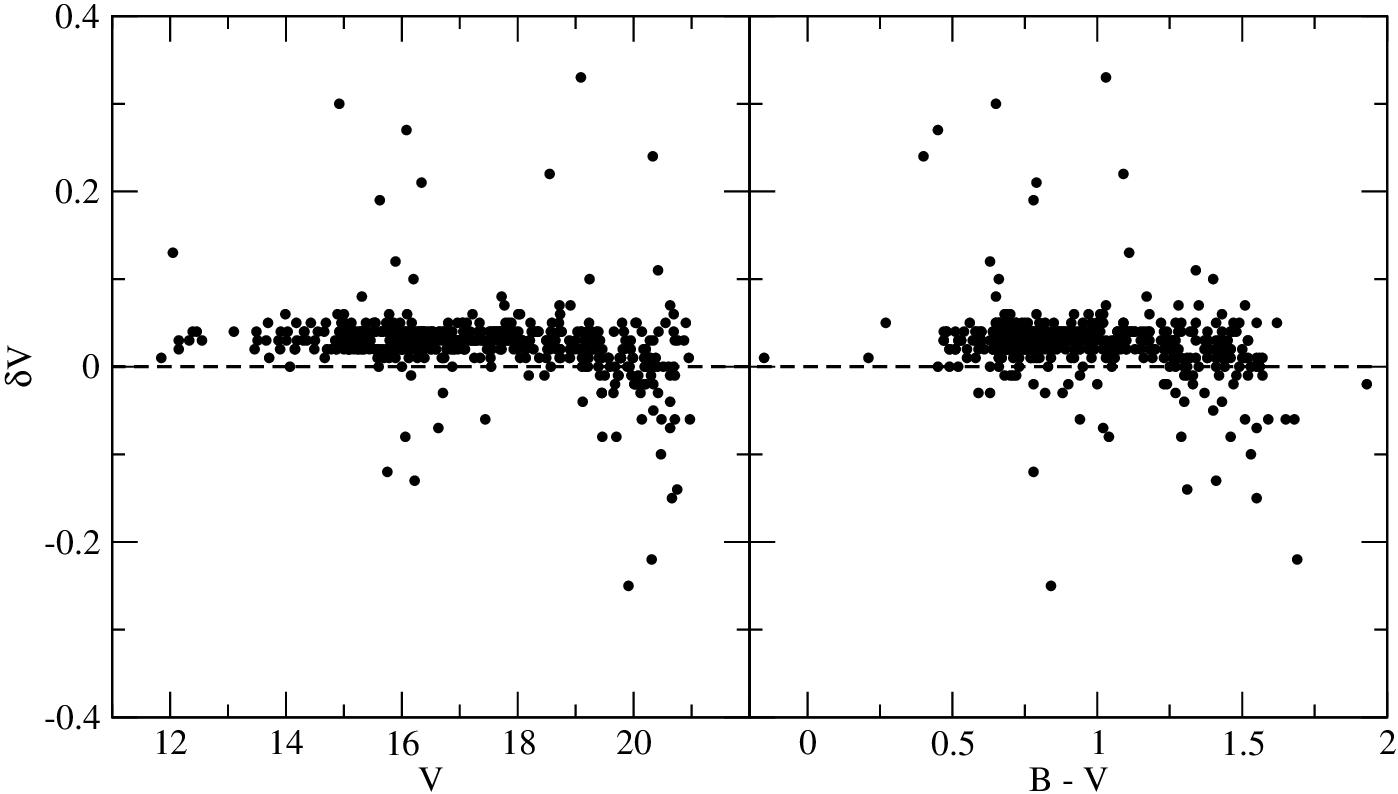}
\caption{A comparison between the V band photometry from 
\cite{ste04} and this paper.  \label{n188compare}}
\end{figure}

This paper's NGC 188 photometry was compared to that of \cite{ste04}
on a star by star basis.  We were able to compare photometry of 533
stars in B, 531 stars in V, 525 stars in I and 304 stars in U.  
%stars 145 and 173 in our list were not matched with a Stetson star
Zero-point offsets of $+0.026\,$mag in U, $+0.013\,$mag in B,
$+0.031\,$mag in V and $+0.044\,$mag in I were found, in the sense
that our photometry was fainter in all filters.  The residuals between
our photometry and \cite{ste04} had no systematic trends with
magnitude or colour. A sample comparison is shown in Figure
\ref{n188compare}, where the differences between our V photometry and
\cite{ste04} are plotted as a function of magnitude and colour.  A
similar comparison with the NGC 188 photometry of \cite{sar99}
determined somewhat smaller zero point offsets of $+0.025\,$mag in U,
$-0.028\,$ mag in B, $-0.002\,$ mag in V and $+0.016\,$mag, in the
sense that our photometry was fainter in U and I, and brighter in B
and V.  Again, the comparison revealed no systematic trends with
magnitude or colour.  It is not clear whose data is best calibrated onto the 
standard system.
Based upon these comparisons it appears that our
photometry is on the \cite{lan92} photometric system to an accuracy of
$\pm 0.03\,$mag.

\begin{figure}
\epsscale{.75}
\plotone{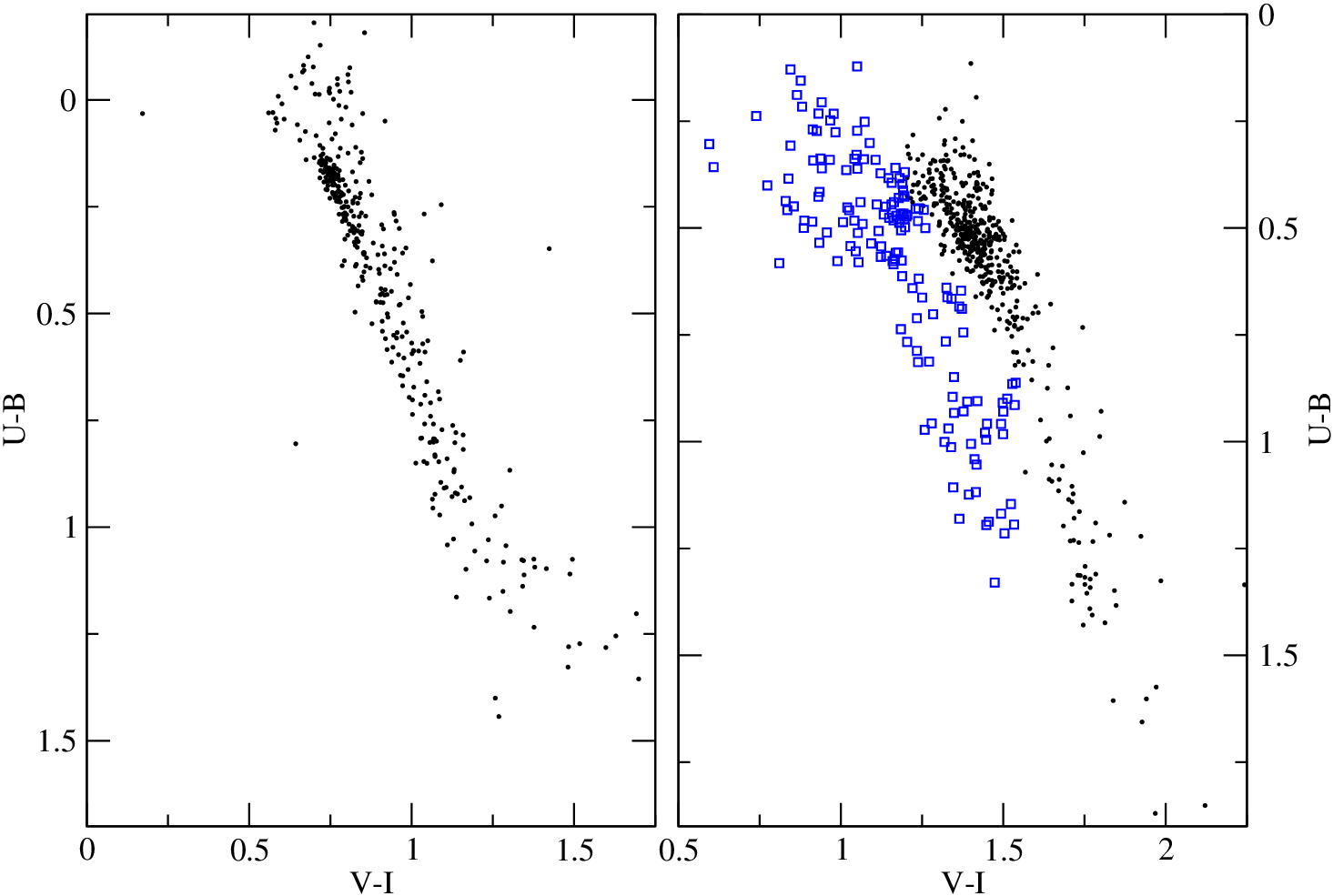}
\caption{NGC 188 (left) and Berkeley 17 (right) $(U-B, V-I)$
Color-Color Diagrams.  Eliminated stars in the Berkeley 17 Color-Color
Diagram are marked with open squares.\label{n188be17color}}
\end{figure}

\begin{figure}
\epsscale{.75}
\plotone{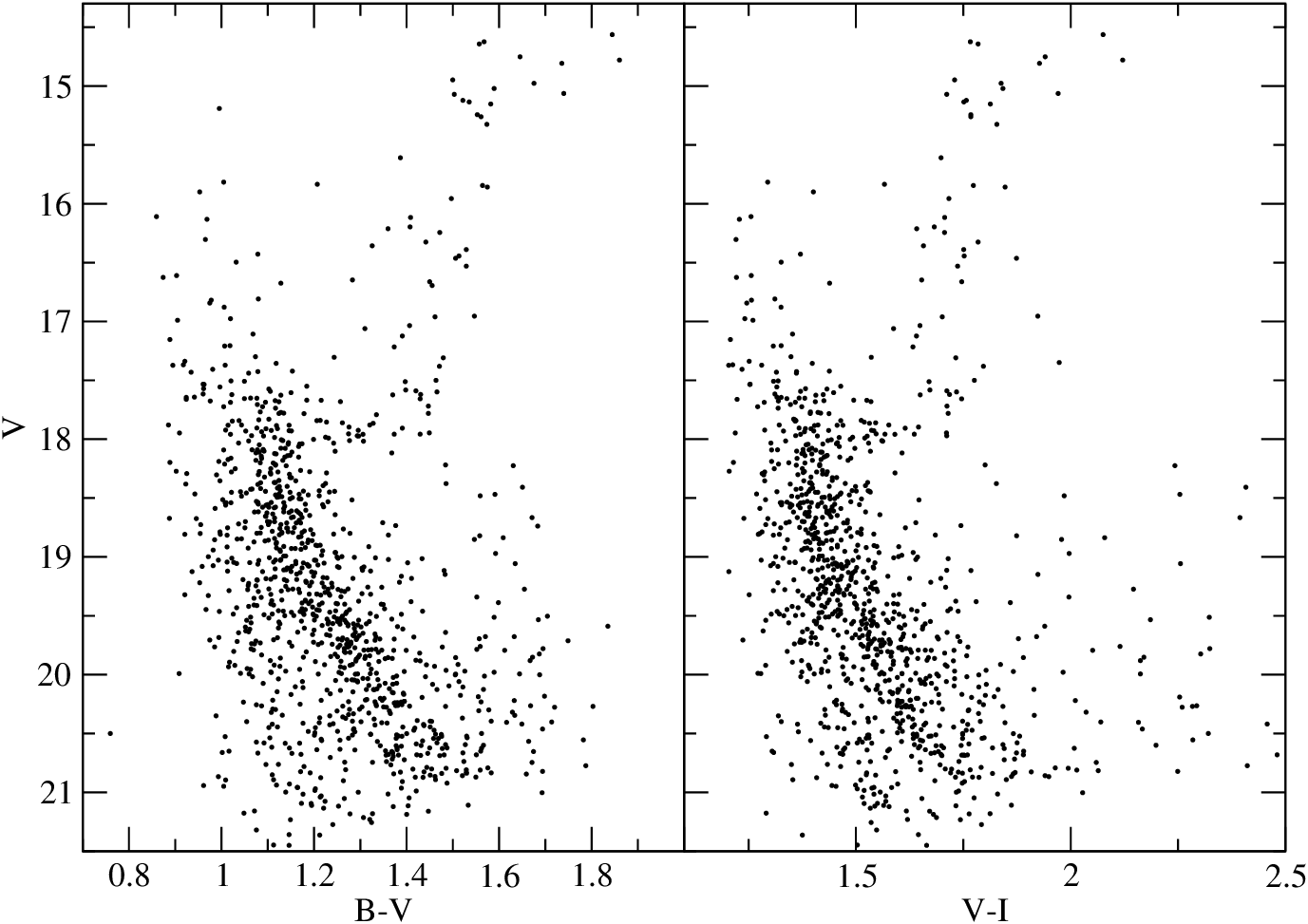}
\caption{Cleaned Berkeley 17 $(V, B-V)$ and $(V, V-I)$
CMDs.\label{cleanbe17cmd}}
\end{figure}

As a result of Berkeley 17's large distance and low Galactic latitude,
its CMD is highly contaminated by field stars.  The CMD consists of a
total of 1239 stars, and extends approximately 4 magnitudes fainter
than the main-sequence turnoff, located at $V\approx 18$.  A highly
contaminated main-sequence can be discerned, along with a
well-populated subgiant branch and a scarcely populated giant branch.
Neither a helium-burning clump not a binary sequence can be discerned.
The $(U-B, V-I)$ colour-colour diagram can be used to estimate
reddening, and thereby remove many foreground stars (which typically
have a lower reddening) from the CMD of Berkeley 17.  The
colour-colour of NGC 188 (which has a low reddening, and few
foreground stars) was compared with the Be 17 colour-colour diagram to
partially eliminate field star contamination (Figure
\ref{n188be17color}).  The cleaned Berkeley 17 CMD contains 1083
stars, and is shown in Figure \ref{cleanbe17cmd}.  Stars which are
deemed to be non-members of Be17 are denoted by a zero in the final
column of Table \ref{be17phottable}.

\begin{figure}
\epsscale{.75}
\plotone{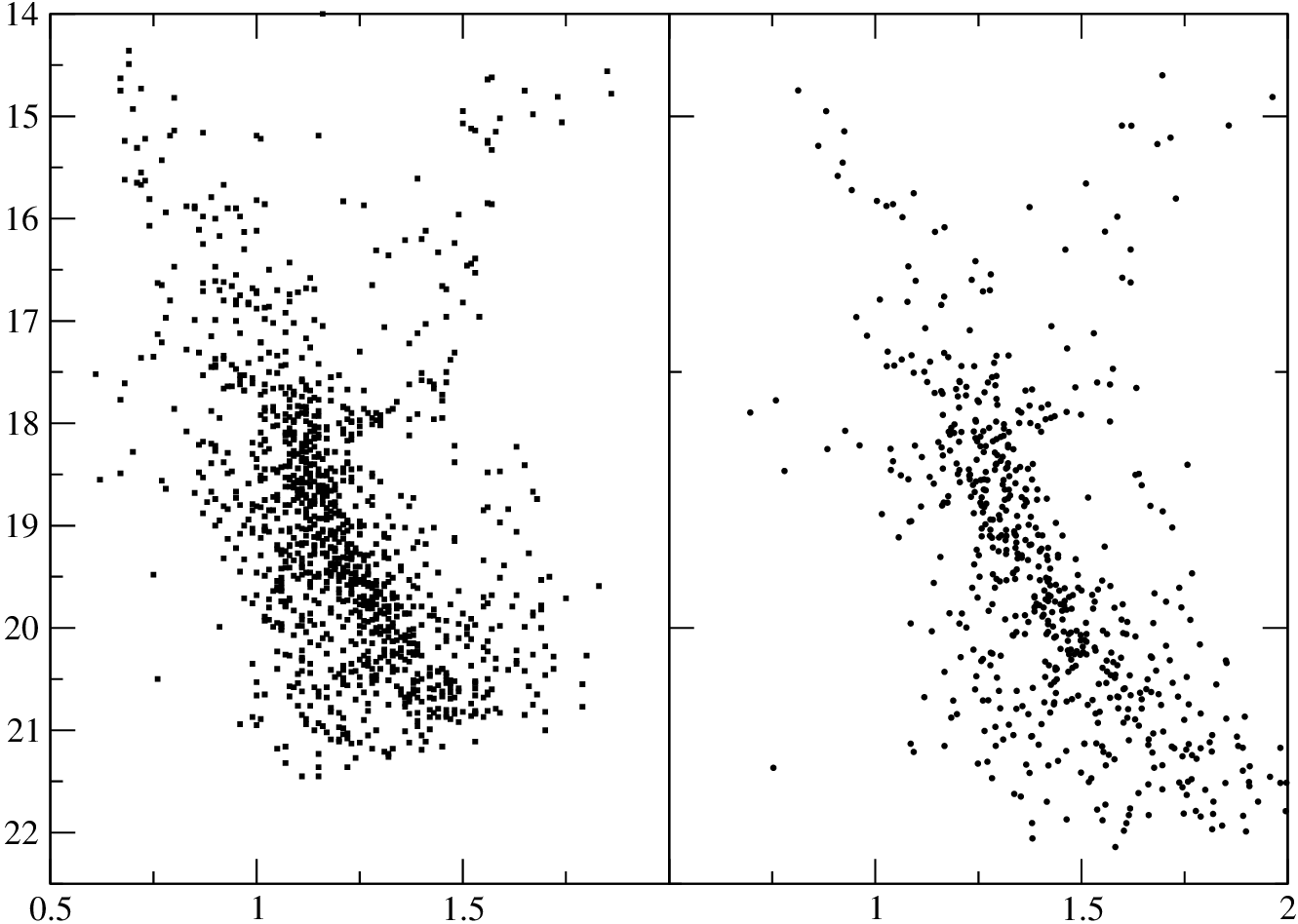}
\caption{Comparison between this paper's Berkeley 17 $(V, B-V)$
photometry shown on the left, and that of Phelps (1997), shown on the
right.\label{be17comp}}
\end{figure}

A comparison between this paper's $(V, B-V)$ photometry (before field
star removal) and that of \cite{phe97} is shown in Figure
\ref{be17comp}.  While the Phelps photometry (which surveyed a
somewhat smaller region) reaches approximately a magnitude deeper,
this paper's photometry displays both a tighter main-sequence and a
more well-defined main-sequence turnoff.  The Phelps photometry is
redder than our photometry, with $\delta (B-V) \sim 0.1\,$mag.  Given
the good agreement of our NGC 188 photometry with the photometry from
\cite{sar99,ste04} we believe that our photometry is on the standard
\cite{lan92} system to within $\pm 0.03\,$mag.

\section{Isochrone Fitting and Age Determinations} \label{section4}

\subsection{Stellar Models and Isochrones}

Stellar evolution tracks were constructed using our stellar
evolution code \citep{cha99,cha01} for masses in the range of 0.5
M$_{\odot}$ to 2.0 M$_{\odot}$ in increments of 0.05 M$_{\odot}$.  The
models were evolved in 2000 time steps from the zero age main-sequence
through the red giant branch.  The models include the diffusion of
helium and heavy elements.  A solar calibrated model was calculated,
yielding $Z_{\odot} = 0.02$ and $Y_{\odot} = 0.275$ (initial
abundances). For other metallicities, ${dY}/{dZ}= 1.5$ was assumed
(corresponding to a primordial helium abundance of $Y_{BBN} = 0.245$).
The solar calibrated mixing length was used for all the models.  The
isochrones were created for metallicities of  [Fe/H] = +0.20,
+0.10, +0.00, --0.12, --0.30, --0.40, --0.60, --0.70, --0.82, and
--1.00.

The transformation from the theoretical log(L) versus log(T$_{eff}$)
plane to observed colors and magnitudes for the constructed isochrones
was completed using two separate color transformations, as this
represents one of the greatest uncertainties of the theoretical
isochrones.  The first transformation utilized the \cite{kur93} model
atmospheres as described in \cite{cha99}, and the second, the
\cite{van03} color tables. The isochrones were constructed in the age
range of 100 Myr to 15 Gyr for each of the
aforementioned metallicities and color transformations, thereby
allowing for considerable flexibility in performing the isochrone fits
to the cluster CMDs.

\subsection{Isochrone Fitting} \label{isofitsection}

One of the most important isochrone input parameters is metallicity,
and therefore, accurate cluster metallicities are crucial to obtaining
reliable age estimates.  We are interested in obtaining precise
relative ages for NGC 188 and Be 17.  For this reason, we elected to
use the metallicity estimates from \cite{fri02}, who conducted an
extensive, homogeneous spectroscopic survey of a large number of old
open clusters, seeking to accurately determine both cluster
metallicities and reddenings.  This paper's clusters, NGC 188 and
Berkeley 17, were both included in their investigation.  As the
\cite{fri02} metallicities for these two clusters were done in a
homogeneous manner, it is likely that the metallicity difference
between the two clusters was robustly measured.  

For relative age
determinations, using precise relative metallicity determinations is
more important than the zero-point of the metallicity scale.  In this
regard, high-resolution spectroscopy of M67
\citep{hob91,tau00} found [Fe/H] values which are about 0.1 dex higher
than \cite{fri02}.  Thus, the \cite{fri02} study may be
underestimating the metallicity of the clusters.  This implies that
our absolute age determinations could be systematically in error, but
does not effect our relative age determinations.  A 0.1 dex increase
in [Fe/H] would systematically decrease our age estimates by approximately 
8\%.  

\cite{fri02} used  Fe I and Fe-peak blends to measure metallicities 
and the H$\beta$
strength of main sequence stars spanning the distance to the cluster
were used to determine reddening values.  \cite{fri02} determined
$[\mathrm{Fe/H}] = -0.10\pm 0.09$ and $\mathrm{E(B-V)} = 0.04\pm 0.04$
for NGC 188.  However, \cite{fri02} note that their H$\beta$ reddening
value is lower than published estimates and adopt $\mathrm{E(B-V)} =
0.08$.  For Be 17, \cite{fri02} found $[\mathrm{Fe/H}] = -0.33\pm
0.12$ and $\mathrm{E(B-V)} = 0.58\pm 0.09$.

In fitting the isochrones to the CMDs of the two clusters, fits are
made to the upper main-sequence, main-sequence turnoff  and sub-giant 
regions.  The
colour of the red giant branch was not used in the fitting process, as
it's location in the theoretical models depends sensitively on
super-adiabatic convection which is not well understood. Similarly,
the colour of the lower main sequence was not used in the fit, as
there are difficulties with the conversion between theoretical
temperatures and colours for cool main sequence stars.  The range of
acceptable isochrone fits is constrained by the requirement of
simultaneous fits in the $(V, B-V)$ and $(V, V-I)$ CMDs.  The
metallicity, reddening, and distance modulus are required to be
identical in both CMDs, assuming $E(V-I) = 1.25 \times [E(B-V)]$
\citep{dea78}.  Metallicities and reddenings are required to be
consistent with those obtained by \cite{fri02}.  Isochrone fit
parameters are summarized in Table \ref{table5}.

\subsection{Discussion}

\subsubsection{NGC 188}

\begin{figure}
\epsscale{.75}
\plotone{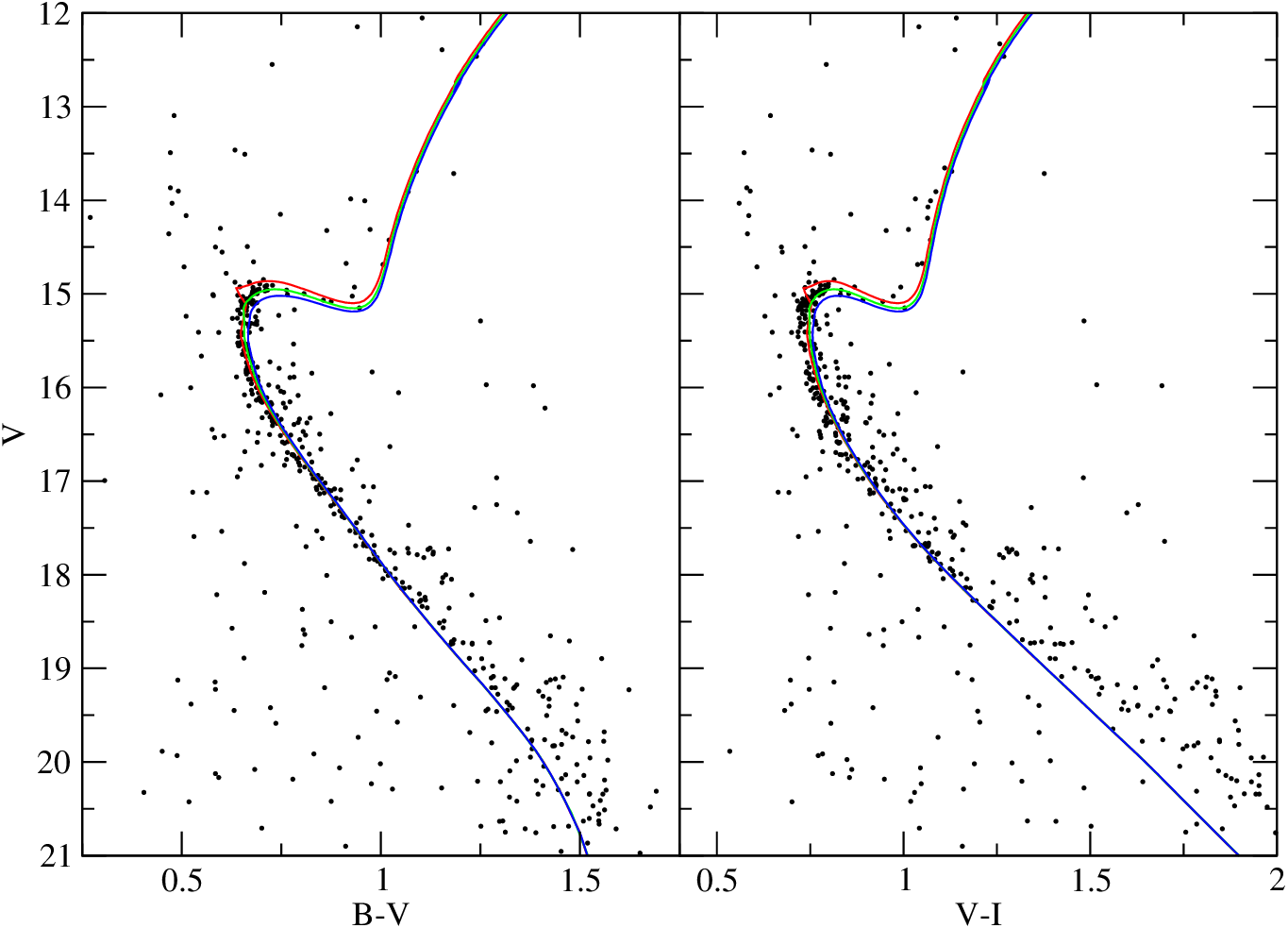}
\caption{Best simultaneous isochrone fits for NGC 188.  Isochrones
constructed with the \cite{van03} colour transformation 
with $[\mathrm{Fe/H}]=-0.12$ for 6.5, 7.0, and 7.5 Gyr are
plotted assuming $\mathrm{E(B - V)} = 0.07$ and $(m - M)_V = 11.35$.
\label{n188isofit}}
\end{figure}

As a result of the high quality of the photometry, and its distinct
CMD morphology NGC 188 is determined to be $7.0 \pm 0.5\,$Gyr old.
The results obtained using the two individual sets of color
transformations are consistent, and an excellent agreement with the
spectroscopic values from \cite{fri02} is also evident.  A sample
isochrone fit is shown in Figure \ref{n188isofit}.  An age of $7.0\pm
0.5\,$Gyr agrees well with the previous age determination of $7.0\pm
0.5\,$Gyr, by \cite{sar99}; $6.8\pm 0.7\,$Gyr by \cite{van04} and
$6.4\,$Gyr by \cite{mic04}.  These studies utilized different photometry
and stellar evolution models.  The agreement between the different studies
suggest that the age of NGC 188 has been robustly determined.

\subsubsection{Berkeley 17}

\begin{figure}
\epsscale{.75}
\plotone{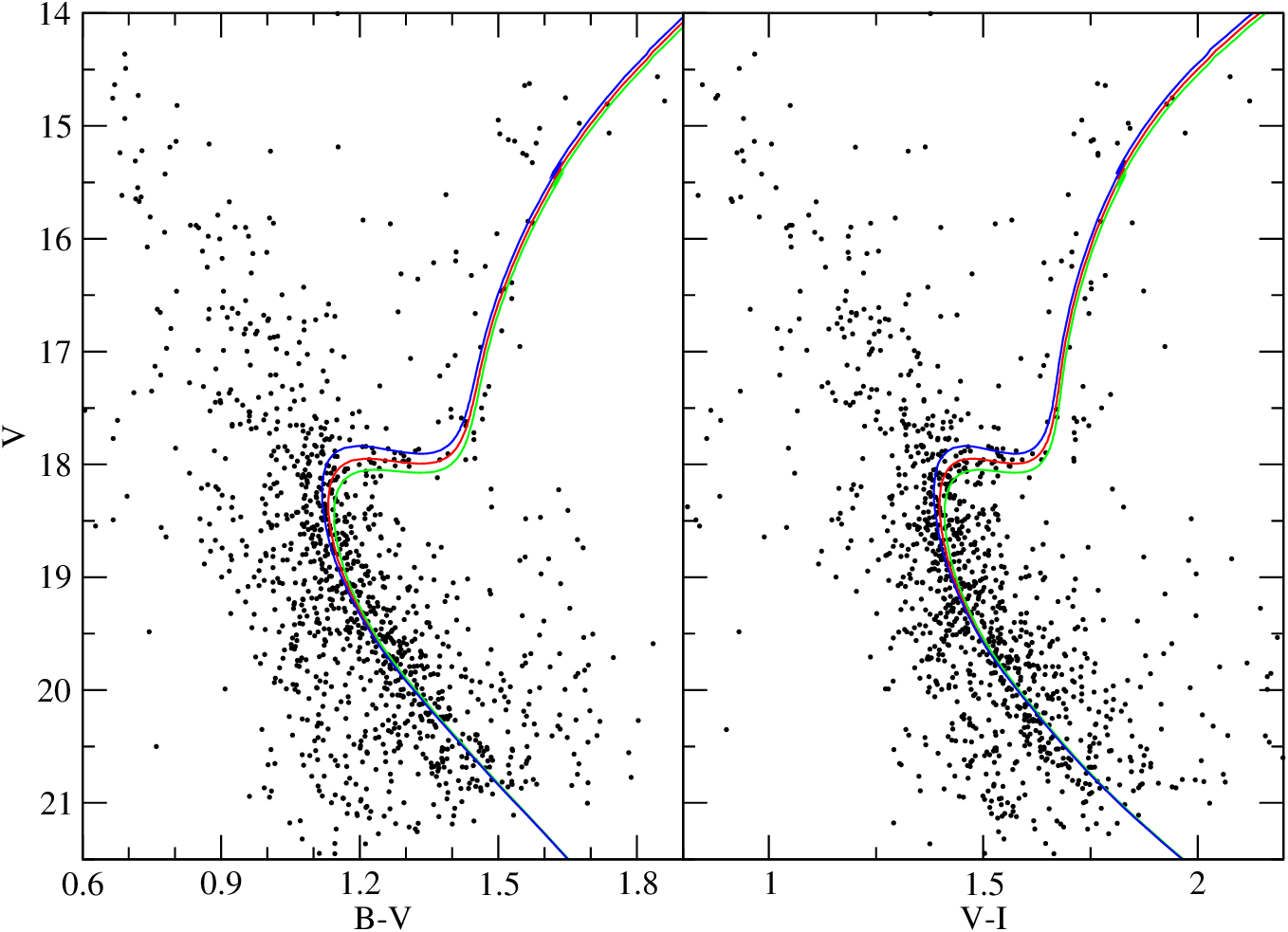}
\caption{Best simultaneous isochrone fits for Berkeley 17.  Isochrones
constructed with the \cite{van03} colour transformation 
with $[\mathrm{Fe/H}]=-0.30$ for 9, 10 and 11 Gyr are
plotted assuming $\mathrm{E(B - V)} = 0.57$ and $(m - M)_V = 14.10$.
\label{be17isofit}}
\end{figure}

The age of Berkeley 17 is found to be 10.0 $\pm$ 1.0 Gyr.  As a result
of the larger reddening uncertainty, greater field star contamination, and 
the less distinct cluster CMD morphology, the established
age for Berkeley 17 has a greater uncertainty than does NGC 188.
However, by using the constraints on reddening and metallicity from
spectroscopy and requiring simultaneous fitting in $(V, B-V)$ and $(V,
V-I)$, reliable results are obtained.  Similar ages are obtained using
the two separate color transformations.  A sample isochrone fit is
shown in Figure \ref{be17isofit}.

Our age for Be 17 is on the low side of the estimates made by \cite{phe97},
who determined an age of 12.5 Gyr based on the \cite{van85}
isochrones, and 10.0--13.0 Gyr based on the \cite{ber94} isochrones.
However, as no accurate spectroscopic study had been performed for
Berkeley 17 at the time, \cite{phe97} allowed for a large range
of metallicities and reddenings.  Additionally, simultaneous fitting
in the $(V, B-V)$, $(V, V-I)$ CMDs was not an important factor in
Phelps' isochrone fitting process, mainly because the isochrones were
not normalized to the Sun either in terms of helium abundance or in
terms of the mixing-length parameter.

On the other hand, \cite{car99a} found an lower estimate of $9.0 \pm
1.0$ Gyr using the \cite{gir00} isochrones on both the \cite{kal94}
and \cite{phe97} photometry.  \cite{car99b}, using the same methodology
as \cite{car99a} made age determinations
of previously well-studied clusters NGC 188 and NGC 6791, for which
age estimates of 6.5 $\pm$ 0.5 Gyr (NGC 188) and 8.5 $\pm$ 0.5 Gyr
(NGC 6791) were determined.  The \cite{car99b} results imply an age
difference between NGC 188 and Berkeley 17 of approximately 2.5 Gyr,
consistent with the results obtained in this paper.  The fact that our
age estimates for Be 17 agree with \cite{phe97} and \cite{car99a} is
surprising, given that our photometry has a significant offset
from the \cite{phe97} photometry (see Figure \ref{be17comp}).

\section{Comparison to Other Clusters} \label{compare}

In order to study the early formation history of the Milky Way, it is
useful to compare the derived ages of NGC 188 and Be17 to other old
open clusters and globular clusters.  As accurate estimates of the
metallicity of a cluster are critical for age determinations, open
clusters were selected from the \cite{fri02} metallicity study which
included [Fe/H] measurements for 39 clusters.  From this list,
clusters were selected which had good quality photometry and a
morphological age index \citep{jan94} greater than 2.0 (corresponding
to an age of approximately 1 Gyr).  The final sample of open clusters
contains 18 clusters.

Globular clusters were selected from the extensive, homogeneous HST
photometric study of globular clusters carried out by \cite{pio02}.
We elected to restrict our sample to relatively metal-rich
($[\mathrm{Fe/H}] > -1.2$) clusters located in the inner part of the
galaxy ($R_{GC} \le 8\,$kpc).  Requiring the globular
clusters to be relatively metal-rich makes the age comparisons to the
more metal-rich open clusters more direct.  The restriction on
Galacto-center distance was made in an attempt to only include
clusters which were formed as part of the initial Galactic collapse,
and not as part of a latter accretion event.  \cite{sal02} found that
globular clusters within a Galactocentric radius of 8 kpc are nearly coeval,
implying that the inner halo globular clusters formed in the initial Galactic
collapse, whereas the outer halo --- where clusters were found to have
a larger range in age --- was likely formed through the accretion of
extra-galactic fragments.  Finally, the globular classification
provided in the recent study by \cite{mac05} was used to exclude
clusters belonging to the young halo (YH), as these are believed to
have been formed in external satellites eventually accreted by the
Milky Way.  Our globular cluster sample contains 13 clusters with B,V
photometry from \cite{pio02}.  For three of these clusters --- 47 Tuc,
NGC 5927, and NGC 6652 --- HST VI photometry was also available.

Given the sensitivity of isochrone fitting to the assumed metallicity, 
it is important that a consistent metallicity scale be used for the 
open and globular clusters.  While numerous studies have attempted
to determine uniform globular cluster \citep{car97,rut97,kraft03} and
open cluster \citep{pia95,twa97,fri02} metallicities, these studies
have generally not overlapped between the two cluster populations.

\cite{rut97} used equivalent width observations to establish a linear
relationship between Ca II triplet equivalent widths and the globular
cluster abundance scale of \cite{car97}.  This lead to a relative
metallicity abundance ranking for a total of 71 globular clusters.
\cite{col04} sought to extend the Ca II triplet abundance scale to the
younger ages and higher metallicities of open clusters.  New
spectra were obtained for a sample of five globular clusters and six
old open clusters, and using globular cluster abundances on the
\cite{car97} scale and open cluster abundances on the \cite{fri02}
scale, a linear relationship between [Fe/H] and Ca II line strength
was found.  A crucial point in their study was to establish the
compatibility of the two metallicity scales, given that the
calibration of two scales differs in two respects: the atmosphere
models utilized, and the effective temperature derivations. 
\cite{col04} conclude that
the systematic offsets between the two scales are comparable to or
smaller than their respective internal uncertainties.  
They believe that the offset resulting from the choice of model
atmospheres is of similar magnitude but opposite sign of that of the
choice of effective temperature scale.  \cite{col04} conclude that
the offset between the two metallicity scales is no more than 0.1 dex.
Globular and open cluster metallicities were thus taken from \cite{rut97}
and \cite{fri02}.

Individual ages for all of the clusters were determined using main
sequence fitting, in the same manner that was done for NGC 188 and Pal
11.  In making these fits, the metallicity of the isochrone was
required to be consistent to within 0.1 dex of the cluster metallicity.
The isochrones were constructed using scaled solar
compositions, while globular clusters are typically enhanced in their
$\alpha$-element (O, Mg, Si, S, and Ca) abundances.  To take this into
account, the relationship between the heavy elements mass fraction
($Z$) in the isochrones and [Fe/H] followed the relationship given by
\cite{chi91} and \cite{cha92}.  In making this correction, globular
cluster $[\alpha/\mathrm{Fe}]$ abundances were assumed to follow the
relationship between $[\alpha/\mathrm{Fe}]$ and [Fe/H] found in
\cite{edv93}.  The $\alpha$-abundances of 47 Tuc were found
to correspond well with this relationship \citep{car04}.  The derived
ages are listed in Table \ref{allages}.

The globular clusters in the sample can be divided up into three
categories based on their location in the Galaxy and metallicity:
thick disk, bulge, and inner halo. The three thick disk globular
clusters, 47 Tuc, NGC 5927, and NGC 6838 are found to have a mean age
of $12.3 \pm 0.4$ Gyr.  The three bulge clusters, NGC 6304, NGC 6624,
and NGC 6637 have a mean age of $12.7 \pm 0.2$ Gyr.  The seven
clusters in the inner halo have an age range of 11.8--14.0 Gyr, with a
mean age of $13.0 \pm 0.9$ Gyr.\footnote{The errors in the mean ages
of the three globular cluster populations represent the standard
deviation of the ages with respect to the mean, and do not include
errors due to isochrone fitting.  The errors in the ages of individual
clusters given in Table \ref{allages} represent relative, rather than
absolute, errors, and thereby, no conflict with cosmological
constraints on the age of the Universe exists.}  Collectively, the
thirteen high-metallicity globular clusters in the sample are found to
be effectively coeval, with an average age of $12.8 \pm 0.7$ Gyr and
with all ages falling within 2$\sigma$ of the mean.  We note that in
general, our globular cluster ages are somewhat older ($\sim 10\%$) than other
studies which attempt to get good absolute ages for globular clusters
\citep{sal02,kra03,lew05}.

The oldest open clusters are found to be NGC
6791, Berkeley 17, and Collinder 261, with ages of $10.0 \pm 1.0$ Gyr,
$10.0 \pm 1.0$ Gyr, and $8.0 \pm 1.0$ Gyr, respectively.  As a result
of its extreme physical parameters --- it is one of the oldest, most
metal-rich, and most populous open clusters --- NGC 6791 has been investigated
in countless photometric and spectroscopic studies.  The age obtained
for NGC 6791 in this study, $10.0 \pm 1.0$ Gyr, is in excellent
agreement with the result found by \cite{sal04}, who determined an age
of $10.2 \pm 1.2$ Gyr.  

\begin{figure}
\epsscale{.75}
\plotone{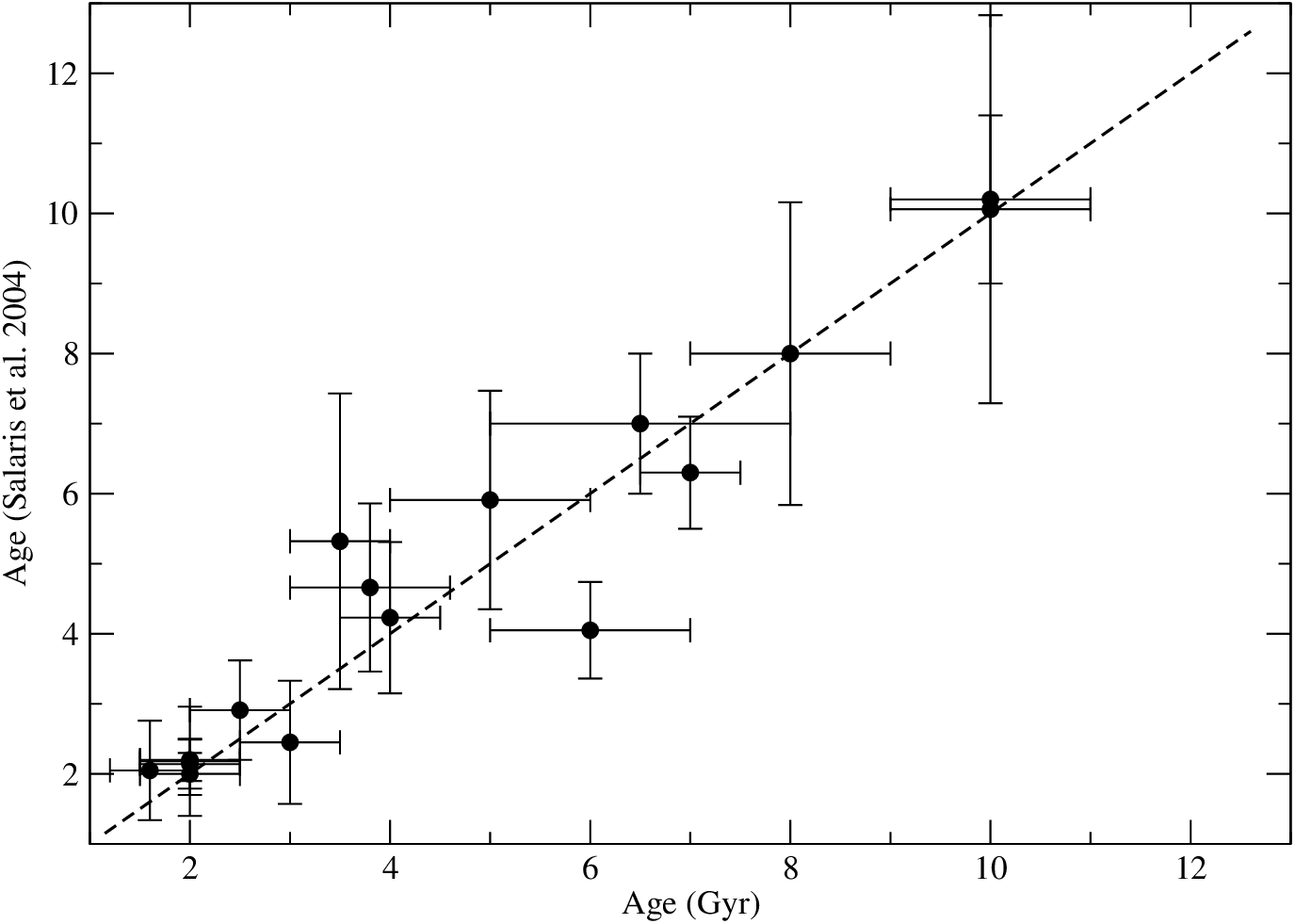}
\caption{Age determinations for the 17 clusters in common between 
this study (horizontal axis) and the ages derived by \cite{sal04}.  
The dashed line has unit slope and illustrates perfect agreement; it is 
not a fit to the data.  The 
only cluster whose age, within the error bars, does not agree
is Be 20 (with an age of 6.0 Gyr in our study).  
\label{agecompare}}
\end{figure}

\cite{sal04} used a combination of main sequence fitting and a new
calibration of the morphological age indicator to determine the age of
71 old open clusters.  The age of NGC 188 from isochrone fitting was
found to be $6.3\pm 0.8\,$Gyr, while the derived age for Be 17 was
$10.1\pm 2.8\,$Gyr.  Our new photometry has allowed us to put a much
tighter constraint on the age of Be 17.  A comparison between the 17
open cluster ages in common between this study and \cite{sal04} is
found in Figure \ref{agecompare}.  In general, our open cluster ages
are found to be consistent, within the error bars, with the ages
obtained in \cite{sal04}.  On average, the error bars in our age
determinations are $\sim 50\%$ smaller than the error bars in the
\cite{sal04} age determinations.

%No sal04 age for NGC 2682

The only cluster for which our age disagrees with the age obtained by
\cite{sal04} is Berkeley 20.  Be 20 is one of the most distant known
open clusters, with a Galactocentric distance of 16.4 kpc. The BVI
photometry for this cluster \cite{dur01} contains significant field
star contamination.  The main-sequence turnoff, the most important
region in the CMD in performing the isochrone fitting, was
poorly-defined.  However, it is evident from the cluster CMD that no
so-called ``hook'' exists in the main-sequence turnoff region, a
characteristic CMD feature of clusters with ages below around 5.0 Gyr.
Thereby, it is unlikely the cluster has an age below 5.0 Gyr, as
suggested by the \cite{sal04} study.  Furthermore, simultaneous
fitting in the $(V, B-V)$ and $(V, V-I)$ CMDs was only possible with
isochrones of around 6 Gyr, for which the reddening value was also
consistent with the value given in the literature ($E(B-V) =
0.14$). Berkeley 20 is thus determined to be $6.0 \pm 1.0$ Gyr old,
which is consistent with the age of 5.0 Gyr obtained from isochrone
fitting by \cite{dur01}

\section{Summary}  \label{section5}

This paper confirms the well-established age of old open cluster NGC
188 at $7.0 \pm 0.5\,$Gyr.  Be 17 is found to be $10.0 \pm 1.0\,$Gry
old.  As a result of the internal consistency of the ages obtained for
the two clusters, the age difference of Be 17 relative to NGC 188 is
highly robust: Be 17 is $3.0 \pm 1.1\,$Gyr older than NGC 188.  These
ages were determined using the metallicity scale of \cite{fri02}.
There are some indications that the \cite{fri02} [Fe/H] measurements
are too low by about 0.1 dex as high resolution spectroscopy studies
of M67 \citep{hob91,tau00} find [Fe/H] values 0.1 dex higher than
\cite{fri02}.  If the \cite{fri02} metallicities are 0.1 dex too low,
then our ages need to be revised downward by about 8\%.

Using the same age determination technique, the \cite{fri02} metallicity
measurements and photometry from the literature, ages of 16 other old,
open clusters were determined.  Be 17 was found to be the oldest open
cluster, with the same age as the well studied cluster NGC 6791.  The
ages of 13 relatively metal-rich globular clusters were determined
using the same methodology as the open cluster ages.  Table
\ref{allages} summarizes the age determinations.  These ages were
determined using isochrone fitting in a self-consistent manner.  This
leads to precise relative ages, but the absolute ages have
considerably larger uncertainties.  We note that the globular cluster
ages we derive are $\sim 8\%$ larger than the accurate absolute ages
found by \cite{kra03}.  This is the same age reduction estimated for
the open clusters, based upon the suggestion that the \cite{fri02}
metallicity values are too low by 0.1 dex.

The globular clusters have ages in the range 11.8 -- 14.0 Gyr, with no
evidence for an age range.  The thick disk globular clusters (the most
metal-rich clusters in our sample) were found to have an average age
of $12.3\pm 0.4\,$Gyr, while the two oldest open clusters in our
sample have an age of $10.0\pm 0.7\,$Gyr.  Thus, the oldest open
clusters in the thin disk are found to be $2.3\pm 0.8\,$Gyr younger than the
thick disk globular clusters.

In contrast, \cite{sal04} find no significant age difference between
the thin and thick disks in the Galaxy.  However, their age
determinations generally had larger error bars than our
determinations, and \cite{sal04} determined the age of two thick disk
globular clusters, while we determine the age of three thick disk
globular clusters.  Combined, these two effects made it difficult for
\cite{sal04} to determine if age differences less than 3 Gyr existed.
In summary, we find that the oldest open clusters imply that the 
thin disk started forming $2.3\pm 0.8\,$Gyr after the formation of the 
thick disk.

\acknowledgments
We would like to thank the referee, whose thoughtful reading of our 
manuscript lead to several improvements.
Research supported in part by a NSF CAREER grant 0094231  to BC.
BC is a Cottrell Scholar of the Research Corporation.

\clearpage

\begin{deluxetable}{l l l c c c}
\tabletypesize{\scriptsize}
\tablecaption{Log of Observations \label{table3}}
\tablewidth{0pt}
\tablehead{
\colhead{Date} & \colhead{Cluster} & \colhead{Exposure (s)} &
\colhead{Filter} & \colhead{Airmass} & \colhead{FWHM (arcsec)}
}
\startdata
    2004 Sep 10&Be 17&3 $\times$ 400&U&1.4&1.8\\
    & &3 $\times$ 90&B&1.3&1.7\\
    & &3 $\times$ 40&V&1.4&1.4\\
    & &3 $\times$ 40&I&1.4&1.4\\

    &Be 17 Off-Field&3 $\times$ 400&U&1.2&1.8\\
    & &3 $\times$ 90&B&1.1&1.8\\
    & &3 $\times$ 40&V&1.2&1.4\\
    & &3 $\times$ 40&I&1.2&1.4\\
 
    2004 Sep 15&NGC 188&3 $\times$ 120&U&1.7&2.5\\
    & &2 $\times$ 1200&U&1.7&2.5\\
    & &3 $\times$ 90&B&1.7&2.0\\
    & &2 $\times$ 900&B&1.7&2.0\\
    & &3 $\times$ 60&V&1.7&2.0\\
    & &2 $\times$ 600&V&1.7&2.0\\
    & &3 $\times$ 30&I&1.7&2.0\\
    & &2 $\times$ 300&I&1.7&2.0\\

    &Be 17&2 $\times$ 1200&U&1.2&1.8\\
    & &2 $\times$ 900&B&1.1&1.8\\
    & &2 $\times$ 600&V&1.3&1.6\\
    & &2 $\times$ 300&I&1.3&1.7\\
 
    2004 Sep 16&Be 17 Off-Field&2 $\times$ 1200&U&1.1&2.0\\
    & &2 $\times$ 900&B&1.1&1.7\\
    & &2 $\times$ 600&V&1.4&1.6\\
    & &2 $\times$ 300&I&1.2&1.5\\

\enddata
\end{deluxetable}

\clearpage

\begin{deluxetable}{r l l l l l l}
\tabletypesize{\scriptsize}
\tablecaption{NGC 188 Photometry File Sample\label{n188phottable}}
\tablewidth{0pt}
\tablehead{
\colhead{Number} &
\colhead{X (pix)} & \colhead{Y (pix)} & \colhead{U (mag)} & \colhead{B
(mag)} & \colhead{V (mag)} & \colhead{I (mag)}
}
\startdata

1 & 921.222&1.588&99.99&19.92&18.74&17.31\\
2 & 265.934&4.543&99.99&21.26&19.83&18.05\\
3 & 62.563&5.192&17.45&17.18&16.39&15.57\\
4 & 12.556&9.095&99.99&20.52&19.17&17.60\\
5 & 63.443&9.635&19.66&18.52&17.28&15.94\\
6 & 405.515&9.803&17.64&17.53&16.83&16.00\\
7 & 350.567&11.787&15.38&15.39&14.78&14.03\\
8 & 314.874&15.181&19.74&18.79&17.72&16.44\\
9 & 634.737&17.929&16.65&16.47&15.81&15.04\\
...&...&...&...&...&...&...\\

\enddata
\end{deluxetable}

\clearpage

\begin{deluxetable}{r l l l l l l c}
\tabletypesize{\scriptsize}
\tablecaption{Berkeley 17 Photometry File Sample\label{be17phottable}}
\tablewidth{0pt}
\tablehead{
\colhead{Number} &
\colhead{X (pix)} & \colhead{Y (pix)} & \colhead{U (mag)} & \colhead{B
(mag)} & \colhead{V (mag)} & \colhead{I (mag)} & \colhead{member?}
}
\startdata

1 & 479.796&1.618&99.99&21.98&20.40&17.79&1\\
2 & 445.481&2.832&99.99&20.69&19.06&16.80&1\\
3 & 677.789&3.111&99.99&21.37&19.97&18.11&1\\
4 & 41.351&5.832&99.99&21.27&20.14&18.57&1\\
5 & 391.842&5.954&20.00&19.13&17.79&99.99&1\\
6 & 692.070&5.973&99.99&20.86&19.59&18.04&1\\
7 & 608.705&6.829&99.99&21.66&20.43&18.79&1\\
8 & 689.438&7.807&20.17&19.92&18.94&17.57&1\\
9 & 56.520&9.221&21.39&20.78&19.61&18.00&1\\
...&...&...&...&...&...&...&...\\

\enddata
\end{deluxetable}

\clearpage

\begin{deluxetable}{l c c c c c}
\tablecaption{Isochrone Fit Parameters \label{table5}}
\tablewidth{0pt}
\tablehead{
\colhead{Cluster} & \colhead{[Fe/H]} & \colhead{E(B-V)} &
\colhead{(m-M)$_V$} & \colhead{Age (Gyr)} & \colhead{Color Transformation}
}
\startdata
 NGC 188& $+0.00$  &0.06 &11.40 &6.5 &VandenBerg \& Clem\\
        & $-0.12$  &0.07 &11.35 &7.0 &VandenBerg \& Clem\\
        & $-0.12$  &0.07 &11.35 &7.0 &Kurucz\\
        & $-0.12$  &0.08 &11.45 &6.5 &Kurucz\\
~~\\
Be 17   &$-0.30$  &0.59 &14.20 & 9  &VandenBerg \& Clem\\
        &$-0.30$  &0.57 &14.10 &10  &VandenBerg \& Clem\\
        &$-0.40$  &0.61 &14.20 &9   &VandenBerg \& Clem\\
        &$-0.40$  &0.60 &14.10 &10  &VandenBerg \& Clem\\
        &$-0.40$  &0.58 &14.10 &10  &Kurucz\\
        &$-0.40$  &0.56 &14.00 &11  &Kurucz\\
\enddata
\end{deluxetable}

\clearpage

\begin{deluxetable}{l c r r r c r c}
\tabletypesize{\scriptsize}
\tablecaption{Cluster Parameters \label{allages}}
\tablewidth{0pt}
\tablehead{
\colhead{Name}&
\colhead{$E(B-V)$}&
\colhead{D}&
\colhead{R$_{GC}$}&
\colhead{Class\tablenotemark{a} } &
\colhead{[Fe/H]} & 
\colhead{Age (Gyr)} &
\colhead{Reference}  
}
\startdata
47 Tuc   &0.04 & 4.5& 7.4& TD &--0.78 &$12.8 \pm 1.0$&1\\
NGC 5904 &0.03 & 7.5& 6.2& IH &--1.12 &$12.0 \pm 1.0$&2\\
NGC 5927 &0.45 & 7.6& 4.5& TD &--0.64 &$12.0 \pm 1.0$&2,3\\
NGC 6171 &0.33 & 6.4& 3.3& IH &--0.95 &$13.5 \pm 1.5$&2\\
NGC 6218 &0.19 & 4.9& 4.5& IH &--1.14 &$14.0 \pm 1.0$&2\\
NGC 6304 &0.53 & 6.0& 2.2& B  &--0.66 &$13.0 \pm 1.0$&2\\
NGC 6362 &0.09 & 7.6& 5.1& IH &--0.99 &$14.0 \pm 1.0$&2\\
NGC 6624 &0.28 & 7.9& 1.2& B  &--0.70 &$12.5 \pm 0.5$&2\\
NGC 6637 &0.16 & 9.1& 1.9& B  &--0.78 &$12.5 \pm 0.5$&2\\
NGC 6652 &0.09 &10.1& 2.8& IH &--0.85 &$11.8 \pm 1.0$&2,4\\
NGC 6712 &0.45 & 6.9& 3.5& IH &--0.94 &$12.5 \pm 0.5$&2\\
NGC 6723 &0.05 & 8.7& 2.6& IH &--0.96 &$13.5 \pm 1.5$&2\\
NGC 6838 &0.25 & 4.0& 6.7& TD &--0.73 &$12.0 \pm 1.0$&2\\
~~\\

NGC 188   &0.08 &1.5 &9.3   &OC &--0.10 &$ 7.0 \pm 0.5 $&5\\
NGC 1193  &0.12 &4.0 &12.0  &OC &--0.51 &$ 4.0 \pm 0.5 $&6\\
NGC 2141  &0.33 &4.1 &12.4  &OC &--0.33 &$ 3.0 \pm 0.5 $&7\\
NGC 2204  &0.08 &4.0 &11.5  &OC &--0.32 &$ 2.0 \pm 0.5 $&8\\
NGC 2243  &0.06 &3.7 &10.8  &OC &--0.49 &$ 3.8 \pm 0.8 $&9\\
NGC 2420  &0.05 &2.3 &10.6  &OC &--0.38 &$ 2.0 \pm 0.5 $&10\\
NGC 2506  &0.07 &3.3 &10.9  &OC &--0.44 &$ 2.0 \pm 0.5 $&11\\
NGC 2682  &0.05 &0.8 &9.05  &OC &--0.15 &$ 5.0 \pm 0.5 $&12\\
NGC 6791  &0.15 &4.2 &8.12  &OC &+0.11  &$10.0 \pm 1.0 $&13\\
NGC 6819  &0.16 &2.3 &8.18  &OC &--0.11 &$ 2.5 \pm 0.5 $&14\\
NGC 6939  &0.50 &1.2 &8.70  &OC &--0.19 &$ 1.6 \pm 0.4 $&15\\
Be 17     &0.58 &2.7 &11.2  &OC &--0.33 &$10.0 \pm 1.0 $&5\\
Be 20     &0.14 &8.4 &16.4  &OC &--0.61 &$ 6.0 \pm 1.0 $&16\\
Be 21     &0.76 &5.0 &13.5  &OC &--0.62 &$ 2.0 \pm 0.5 $&17\\
Be 31     &0.20 &3.8 &12.0  &OC &--0.40 &$ 3.5 \pm 0.5 $&18\\
Be 32     &0.15 &3.1 &11.4  &OC &--0.50 &$ 5.0 \pm 1.0 $&19\\
Be 39     &0.12 &3.8 &11.5  &OC &--0.26 &$ 6.5 \pm 1.5 $&8\\
Cr 261    &0.33 &2.4 &7.5   &OC &--0.16 &$ 8.0 \pm 1.0 $&20\\

\enddata

\tablenotetext{a}{Classification:  B Bulge globular cluster;
TD Thick Disk globular cluster; IH Inner Halo globular cluster;
OC Open Cluster}
 
\tablerefs{Galactic Coordinates, reddenings and distances for globular
clusters are from Harris (1996); Galactic Coordinates, reddenings and
distances for open clusters are from \cite{fri02}; Sources for
photometry: (1) Kaluzny et al.\ (1998); (2) Piotto et al.~(2002); (3)
Feltzing \& Gilmore (2000); (4) Chaboyer et al.~(2000); (5) This
paper; (6) Kaluzny (1988); ((7) Carraro et al. (2001);~(8) Kassis et
al.~(1997); (9) Bergbusch et (al.~(1991); (10) Anthony-Twarog et
al.~(1990); (11) Kim et (al.~(2001); (12) Sandquist (2004); (13)
Stetson et al.~(2003); (14) (Kalirai et al.~(2001); (15) Andreuzzi et
al.~(2004); (16) Durgapal et (al.~(2001); (17) Tosi et al.~(1998);
(18) Guetter (1993); (19) (Kaluzny \& Mazur (1991); 20) Gozzoli et
al.~(1996); All OC photometry (obtained from WEBDA, maintained by
Jean-Claude Mermilliod (http://obswww.unige.ch/webda/navigation.html)}

\end{deluxetable}

\clearpage

\clearpage

\clearpage

\clearpage

\clearpage

\clearpage

\clearpage

\clearpage

\clearpage

\begin{thebibliography}{}

\bibitem[Andreuzzi et al.(2004)]{and04}Andreuzzi, G., Bragaglia, A.,
Tosi, M., \& Marconi, G.  2004, MNRAS, 348, 297

\bibitem[Anthony-Twarog et al.(1990)]{ant90}Anthony-Twarog, B. J.,
Kaluzny, J., Shara, M. M., \& Twarog, B. A.  1990, AJ, 99, 5

\bibitem[Bergbusch et al.(1991)]{ber91}Bergbusch, P. A., VandenBerg,
D. A., \& Infante, L.  1991, AJ, 101, 6

\bibitem[Bertelli et al.(1994)]{ber94} Bertelli, G., Bressan, A.,
Chiosi, C., Fagotto, F., \& Nasi, E.  1994, A\&AS, 106, 275

\bibitem[Caputo et al.(1990)]{cap90} Caputo, F., Chieffi, A.,
Castellani, V., Collados, M., Marinez Roger, C., \& Paez, E. 1990, AJ,
99, 261

\bibitem[Carraro et al.(1999a)]{car99a} Carraro, G., Girardi, L., \&
Chiosi, C. 1999a, MNRAS, 309, 430

\bibitem[Carraro et al.(1999b)]{car99b} Carraro, G., Vallenari, A.,
Girardi, L., \& Richichi, A. 1999b, A\&A, 343, 825

\bibitem[Carraro et al.(2001)]{car01}Carraro, G., Hassan, S. M.,
Ortolani, S., \& Vallenari, A.  2001, A\&A, 372, 879

\bibitem[Carretta \& Gratton(1997)]{car97}Carretta, E., \& Gratton,
R. G.  1997, A\&ASS, 121, 95

\bibitem[Carretta et al.(2004)]{car04}Carretta, E., Gratton, R. G.,
Bragaglia, A., Bonifacio, P., \& Pasquini, L.  2004, A\&A, 416, 925

\bibitem[Chaboyer et al.(1992)]{cha92}Chaboyer, B., Sarajedini, A., \&
Demarque, P.  1992, ApJ, 394, 515

\bibitem[Chaboyer et al.(1999)]{cha99} Chaboyer, B., Green, E. M., \&
Liebert, J. 1999, AJ, 117, 1360

\bibitem[Chaboyer et al.(2000)]{cha00}Chaboyer, B., Sarajedini, A., \&
Armandroff, T. E.  2000, AJ, 120, 3120

\bibitem[Chaboyer et al.(2001)]{cha01} Chaboyer, B., Fenton, W. H.,
Nelan, J. E., Patnaude, D. J., \& Simon, F.E.  2001, ApJ, 562, 521

\bibitem[Chieffi et al.(1991)]{chi91}Chieffi, A. Straniero, O., \&
Salaris, M.  1991, in Janes, K., ed, \emph{The Formation and Evolution
of Star Clusters (ASP Conf.\ Ser.\ 13)} (ASP: San Francisco), 219

\bibitem[Cole et al.(2004)]{col04}Cole, A. A., Smecker-Hane, T. A.,
Tolstoy, E., Bosler, T. L., \& Gallagher, J. S.  2004, MNRAS, 347, 367

\bibitem[Dean et al.(1978)]{dea78} Dean, J. F., Warren, P. R., \&
Cousins, A. W. J. 1978, MNRAS, 183, 569

\bibitem[Durgapal et al.(2001)]{dur01}Durgapal, A. K., Pandey, A. K.,
\& Mohan, V.  2001, A\&A, 372, 71

\bibitem[Edvardsson et al.(1993)]{edv93}Edvardsson, B., Andersen, J.,
Gustafsson, B., Lambert, D. L., Nissen, P. E., \& Tomkin, J.  1993,
A\&A, 275, 101

\bibitem[Feltzing \& Gilmore(2000)]{fel00}Feltzing, S., \& Gilmore, G.
2000, A\&A, 355, 949

\bibitem[Friel et al.(2002)]{fri02} Friel, E. D., Janes, K. A.,
Tavarez, M., Scott, J., Katsanis, R., Lotz, J., Hong, L., \& Miller,
N. 2002, AJ, 124, 2693

\bibitem[Girardi et al.(2000)]{gir00} Girardi, L., Bressan, A.,
Bertelli, G., \& Chiosi, C.  2000, A\&AS, 141, 371

\bibitem[Gozzoli et al.(1996)]{goz96}Gozzoli, E., Tosi, M., Marconi,
G., \& Bragaglia, A.  1996, MNRAS, 283, 66

\bibitem[Guetter(1993)]{gue93}Guetter, H. H.  1993, AJ, 106, 220

\bibitem[Harris(1996)]{har96}Harris, W.E. 1996, AJ, 112, 1487
(February 2003 revision)

\bibitem[Hobbs \& Thorburn(1991)]{hob91} Hobbs, L.~M., \& 
Thorburn, J.~A.\ 1991, \aj, 102, 1070 

\bibitem[Janes \& Phelps(1994)]{jan94} Janes, K. A., \& Phelps,
R. L. 1994, AJ, 108, 1773

\bibitem[Kalirai et al.(2001)]{kali01}Kalirai, J. S., Richer, H. B.,
Fahlman, G. G., Cuillandre, J.-Ch., Ventura, P., D'Antona, F., Bertin,
E., Marconi, G., \& Durrell, P. R.  2001, AJ, 122, 266

\bibitem[Kaluzny(1988)]{kal88}Kaluzny, J.  1988, AcA, 38, 339

\bibitem[Kaluzny(1994)]{kal94} Kaluzny, J. 1994, AcA, 44, 247

\bibitem[Kaluzny \& Mazur(1991)]{kal91}Kaluzny, J., \& Mazur, B.
1991, AcA, 41, 167

\bibitem[Kaluzny et al.(1998)]{kal98} Kaluzny, J., Wysocka, 
A., Stanek, K.~Z., \& Krzeminski, W.\ 1998, Acta Astronomica, 48, 439 

\bibitem[Kassis et al.(1997)]{kas97}Kassis, M., Janes, K. A., Friel,
E. D., \& Phelps, R. L.  1997, AJ, 113, 1723

\bibitem[Kim et al.(2001)]{kim01}Kim, S. L., Chun, M.-Y., Park, B.-G.,
Kim, S. C., Lee, S.-H., Lee, M. G., Ann, H. B., Sung, H., Jeon, Y.-B.,
\& Yuk, I.-S.  2001, AcA, 51, 49

\bibitem[Krauss \& Chaboyer(2003)]{kra03} Krauss, L. M., \& Chaboyer,
B.  2003, Science, 299, 65

\bibitem[Kraft \& Ivans(2003)]{kraft03} Kraft, R.~P., \& Ivans, 
I.~I.\ 2003, \pasp, 115, 143 

\bibitem[Kurucz(1993)]{kur93}Kurucz, R.L.  1993, CD-ROM No. 13 (Cambridge:
Smithsonian Astrophysical Observatory)

\bibitem[Landolt(1992)]{lan92} Landolt, A. U.  1992, AJ, 104, 340

\bibitem[Lewis et al.(2005)]{lew05} Lewis, M.S., Liu, W.M., Paust, N.E.Q,
\& Chaboyer, B.\ 2005, \aj, submitted

\bibitem[Liu \& Chaboyer(2000)]{liu00} Liu, W. M., \& Chaboyer,
B. 2000, ApJ, 544, 818

\bibitem[Mackey \& van den Bergh(2005)]{mac05}Mackey, A. D., \& van
den Bergh, S.  2005, MNRAS, 360, 631

\bibitem[Michaud et al.(2004)]{mic04} Michaud, G.,
Richard, O., Richer, J., \& VandenBerg, D.~A.\ 2004, \apj, 606, 452

\bibitem[Phelps et al.(1994)]{phe94} Phelps, R. L., Janes, K. A., \&
Montgomery, K. A. 1994, AJ, 107, 1079

\bibitem[Phelps(1997)]{phe97} Phelps, R. L. 1997, ApJ, 483, 826

\bibitem[Piatti et al.(1995)]{pia95}Piatti, A. E., Claria, J. J., \&
Abadi, M. G.  1995, AJ, 110, 2813

\bibitem[Piotto et al.(2002)]{pio02}Piotto, G., King, I. R.,
Djorgovski, S. G., Sosin, C., Zoccali, M., Saviane, I., De Angeli, F.,
Riello, M., Recio Blanco, A., Rich, R. M., Meylan, G., \& Renzini, A.
2002, A\&A, 391, 945

\bibitem[Platais et al.(2003)]{pla03} Platais, I., Kozhurina-Platais,
V., Mathieu, R. D., Girard, T. M., \& van Altena, W. F.\  2003, AJ,
126, 2922

\bibitem[Rutledge et al.(1997)]{rut97}Rutledge, G. A., Hesser, J. E.,
\& Stetson, P. B.  1997, PASP, 107, 907

\bibitem[Salaris \& Weiss(2002)]{sal02}Salaris, M., \& Weiss, A.
2002, A\&A, 388, 492

\bibitem[Salaris et al.(2004)]{sal04} Salaris, M., Weiss, A., \&
Percival, S. M.  2004, A\&A, 414, 163

\bibitem[Sandage(1962)]{san62} Sandage, A.\ 1962, ApJ, 135, 333

\bibitem[Sandquist(2004)]{san04}Sandquist, E. L.\  2004, MNRAS, 347, 101

\bibitem[Sarajedini et al.(1999)]{sar99} Sarajedini, A., von Hippel,
T., Kozhurina-Platais, V., \& Demarque, P. 1999, AJ, 118, 2894

\bibitem[Scott et al.(1995)]{sco95} Scott, J. E., Friel, E. D., \&
Janes, K. E.  1995, AJ, 109, 1706

\bibitem[Setteducati \& Waever(1962)]{set62} Setteducati, A. F., \&
Waever, M. F. 1962, Newly Found Stellar Clusters (Berkeley: Radio
Astronomy Laboratory, University of California, Berkeley)

\bibitem[Stetson(1992)]{ste92} Stetson, P. B. 1992, ASP Conf. Ser. 25:
Astronomical Data Analysis Software and Systems I, 1, 297

\bibitem[Stetson(2000)]{ste00} Stetson, P.~B.\ 2000, \pasp, 
112, 925 

\bibitem[Stetson et al.(2003)]{ste03} Stetson, P.~B., Bruntt, 
H., \& Grundahl, F.\ 2003, \pasp, 115, 413 

\bibitem[Stetson et al.(2004)]{ste04} Stetson, P.~B., McClure, R.~D.,
\& VandenBerg, D.~A.\ 2004, \pasp, 116, 1012

\bibitem[Tautvai{\v s}iene et al.(2000)]{tau00} Tautvai{\v 
s}iene, G., Edvardsson, B., Tuominen, I., \& Ilyin, I.\ 2000, \aap, 360, 
499 

\bibitem[Tosi et al.(1998)]{tos98}Tosi, M., Pulone, L., Marconi, G.,
\& Bragaglia, A.  1998, MNRAS, 299, 834

\bibitem[Twarog \& Anthony-Twarog(1989)]{twa89} Twarog, B.A., \&
Anthony-Twarog, B.J. 1989, AJ, 97, 759

\bibitem[Twarog et al.(1997)]{twa97}Twarog, B. A., Ashman, K. M., \&
Anthony-Twarog, B. J.  1997, AJ, 114, 2556

\bibitem[VandenBerg(1985)]{van85} VandenBerg, D. A.  1985, ApJS, 58,
711

\bibitem[VandenBerg \& Stetson(2004)]{van04} VandenBerg, D.~A., \&
Stetson, P.~B.\ 2004, \pasp, 116, 997

\bibitem[VandenBerg \& Clem(2003)]{van03} VandenBerg, D. A., \& Clem,
J. L. 2003, AJ, 126, 778

\bibitem[von Hippel \& Sarajedini(1998)]{von98} von Hippel, T., \&
Sarajedini, A.  1998, AJ, 116, 1789

\end{thebibliography}
\end{document}